\documentclass[aps,preprint,showpacs]{revtex4}
\usepackage{epsfig}
\usepackage{amsmath}
\usepackage{amsfonts}
\usepackage{bm}
\usepackage{color}

\newcommand{\bfx}{{\bf x}}

\newcommand{\bfk}{{\bf k}}

\newcommand{\beps}{{\mbox{\boldmath$\epsilon$}}}

\begin{document}
\title[Parity nonconservation in resonance recombination]
{Parity nonconservation effect in resonance recombination of
polarized electrons with heavy hydrogenlike ions}
\date{\today}

\author{A.~V.~Maiorova,$^{1}$ V.~M.~Shabaev,$^{1}$ A.~V.~Volotka,$^{1, 2}$ V.~A.~Zaytsev,$^{1}$ G.~Plunien,$^{2}$ T.~St\"ohlker$^{3, 4, 5}$}
\affiliation {$^{1}$Department of Physics, St.Petersburg State
University, Ulianovskaya 1, Petrodvorets, St.Petersburg 198504, Russia\\
$^2$Institut f\"ur Theoretische Physik, Technische Universit\"at
Dresden, Mommsenstrasse 13, D-01062 Dresden, Germany\\
$^{3}$Gesellschaft f\"{u}r Schwerionenforschung,
Planckstrasse 1, D-64291 Darmstadt, Germany\\
$^4$Physikalisches Institut, Universit\"at Heidelberg,
Philosophenweg 12, D-69120 Heidelberg, Germany\\
$^5$Helmholtz-Institut Jena, D-07743 Jena, Germany}
\begin{abstract}
Parity nonconservation (PNC) effect in recombination of a polarized electron
with a heavy H-like ion in case of resonance with a doubly excited state of
the corresponding He-like ion is studied. It is assumed that photons of
the energy corresponding
to the one-photon decay  of the doubly excited state into the $2^1S_0$
or the $2^3P_0$ state are detected at a given angle with respect to
the incident electron momentum. Calculations are performed for heliumlike
thorium ($Z = 90$) and gadolinium ($Z = 64$), where the  $2^1S_0$ and $2^3P_0$
levels are near to cross and, therefore, the PNC effect is strongly enhanced.

\end{abstract}

\maketitle

\section{Introduction}
Parity nonconservation (PNC) effects caused by the weak neutral-current interaction
were extensively studied in neutral atomic systems
\cite{khr91,khr04,gin04}. Recent progress on the theory
of the PNC effects in neutral atoms is mainly related to
evaluation of the  QED correction and a signifcant improvement
of the accuracy of the electron-correlation contribution
(see Refs. \cite{sha05,por09} and references therein).
The accuracy of the theoretical predictions for the PNC effects
in neutral atoms is mainly  limited by an uncertainty
of the electron-correlation contribution.
In contrast to that,
in highly charged few-electron ions the electron-correlation contribution,
being suppressed by a factor $1/Z$ ($Z$ is the nuclear charge number),
can be calculated to a very high accuracy employing the $1/Z$ perturbation
theory. This provides very good prospects for investigations of the PNC
effects with heavy few-electron ions.

PNC experiments with highly charged ions were first
discussed in Ref. \cite{gor74}, where it was proposed
to use the close opposite-parity levels in He-like ions at $Z \sim
6$ and  $Z \sim 29$. Later,
the PNC effects with heavy ions were considered by a number of authors
\cite{sch89,opp91,kar92,dun96,zol97,lab01,pin93,gri05,lab07,mai09,sha10,fer10,fer11}.
Most of these studies exploited the near-degeneracy of
the $2^1S_0$ and $2^3P_0$ levels in He-like ions at $Z \sim 64$
and $Z \sim 90$ , where the PNC effect is strongly enhanced
(see, e.g., Refs. \cite{sch89,lab01,lab07,mai09,sha10,fer10,fer11}).
In particular, in our recent investigation \cite{mai09} we
evaluated the PNC effect on the cross section of the radiative
recombination of an electron into the $2^1S_0$ and $2^3P_0$ states
of He-like ions for two experimental scenarios.
In the first scenario, the incident electron
is polarized, while the H-like ion is unpolarized, and the photon
polarization is not detected.
In the second one, linearly polarized photons are detected in an
experiment with unpolarized electrons and ions. The numerical
results for both scenarios were obtained and optimum cases in
which the effect is most pronounced were found.

In the present paper, we study the PNC effect on
recombination of a polarized electron with
unpolarized hydrogenlike  thorium ($Z = 90$) and gadolinium ($Z = 64$)
ions in case of resonance with a doubly excited state
of the He-like ion. We consider a scenario in which
photons of the energy equal to the difference between the
initial energy (the ground state energy of the H-like ion
and the incident electron energy) and the energy of the He-like ion
in the $2^1S_0$ or the $2^3P_0$ state are detected at a given angle.
As the intermediate doubly excited states, we consider
$(2s_{1/2}2p_{1/2})_{1}$, $(2s_{1/2}2p_{3/2})_{1}$,
$(2p_{1/2}2p_{3/2})_{1}$,  $(2s_{1/2}3p_{1/2})_{1}$, and
 $(2s_{1/2}3p_{3/2})_{1}$  states,
where the PNC effect on the dielectronic recombination cross section
should be most pronounced.
This is due to the fact that for such states we can choose the magnetic
dipole (M1) transition to the $2^1S_0 \equiv (1s_{1/2}2s_{1/2})_{0}$
or the $2^3P_0 \equiv (1s_{1/2}2p_{1/2})_{0}$ state,
while the PNC-mixing channel is the
electric dipole (E1) transition.
The values of the photon emission angle with respect to the incident
electron momentum, which correspond to the maximum PNC effect,
are evaluated.

Relativistic units ($\hbar=c=1$) and the Heaviside charge unit
($\alpha = e^2/(4\pi)$, $e<0$) are used throughout the paper.

\section{Basic formulas}\label{th}

We consider recombination of an electron having
asymptotic four-momentum $p_i =
(\varepsilon_{i},\mathbf{p_i})$ and polarization $\mu_i$
with a heavy H-like ion in the ground $1s$ state at $Z\approx 90$
or $Z\approx 64$.  We choose the
incident electron energy $\varepsilon_{i}$  to get
the resonance with a doubly excited state $d$ of the He-like ion,
$E_{1s}+\varepsilon_{i}= E_d$,
and assume that photons of the energy corresponding to the decay of this
state to the $2^1S_0$ or the $2^3P_0$ state are detected
in experiment. In the resonance approximation,
the cross section of the process
is the sum of the dielectronic recombination (DR) and radiative
recombination (RR) cross sections and an interference term.
In the scenario we consider
the photon polarization is not measured. Then, neglecting
the weak interaction,  the differential cross section
of the one-photon recombination into the $2^1S_0$ or the $2^3P_0$
final state, $\sigma\equiv d\sigma/d\Omega$, is given by \cite{kar92a,sha94}
\begin{eqnarray} \label{cross}
\sigma=
\frac{(2\pi)^{4}}{{v_{i}}}\mathbf{k_f^2}\sum_{\beps_f}\left|\sum_{M_d}\tau_{\gamma_f,f;d}
\frac{1}{E_i-E_d+i\Gamma_d/2}
\langle\Psi_d|I|\Psi_i\rangle+\tau_{\gamma_f,f;i}
\right|^{2}\,.
\end{eqnarray}
Here $i$, $d$, and $f$ are the
initial, intermediate, and final states, respectively.
 $E_i = E_{1s}+\varepsilon_{i} $ is the
energy of the initial state, $E_d$ and $\Gamma_d$ are the energy
and the width of the intermediate doubly excited state $d$, $I$ is
the operator of the interelectronic interaction,
 $\tau_{\gamma_f,f;i}$ is the RR amplitude into the
final state $f$,
 $\tau_{\gamma_f,f;d}$ is the transition amplitude from
the intermediate state $d$ to the final state,
$\mathbf{k_f} $ is the photon momentum,
and $v_i$ is the incident electron velocity.

For heavy few-electron ions we can generally use the one-electron
approximation for  the wave functions,  since the
interelectronic-interaction effects are suppressed by a factor
$1/Z$, compared to the interaction of the electrons with Coulomb field of the
nucleus. Then the wave function of the initial state is
\begin{eqnarray}
\label{wf_init} \Psi_{i}(\mathbf{x_{1}},\mathbf{x_{2}}) =
\frac{1}{\sqrt{2}}\sum_{\mathcal{P}}(-1)^{\mathcal{P}}\mathcal{P}\psi_{1s}(\mathbf{x_{1}})\psi_{p_{i}\mu_{i}}(\mathbf{x_{2}})\,,
\end{eqnarray}
where $\psi_{1s}(\mathbf{x})$ is the one-electron 1s Dirac wave
function, $ \psi_{p_{i}\mu_{i}}$ is the incident electron Dirac
wave function, $(-1)^{\mathcal{P}}$ is the parity of the
permutation, and $\mathcal{P}$ is the permutation operator.
Assuming that the electron momentum $\mathbf{p_i}$ is directed
along the quantization axis (we consider the process in the ion
rest frame), the wave function of the incident electron can be
expanded as
\begin{eqnarray}
\label{electron_wave_decompose}
\psi_{p_{i}\mu_{i}}(\mathbf{x})=
\frac{1}{\sqrt{4\pi}} \, \frac{1}{\sqrt{p_{i}\varepsilon_{i}}} \,
\sum_{\kappa} \mathrm{i}^{l}\exp(i\Delta_{\kappa}) \, \sqrt{2l+1}
\,
 C_{l0, \,\frac{1}{2}\mu_{i}}^{j\mu_{i}}
\psi_{\varepsilon_{i} \kappa \mu_{i}}(\mathbf{x}) \, .
\end{eqnarray}
where  $\Delta_{\kappa}$ is the Coulomb phase
shift,  $C_{l0, \,\frac{1}{2}\mu_{i}}^{j\mu_{i}}$ is the Clebsch-Gordan coefficient,
and $ \psi_{\varepsilon_{i} \kappa \mu_{i}}(\mathbf{x})$
is the partial
electron wave with the Dirac quantum number
$\kappa=(-1)^{j+l+1/2}(j+1/2)$  determined by angular momentum
\emph{j} and parity of the state \emph{l}.

If we neglect
the weak electron-nucleus interaction,
the wave functions of the \emph{d} and \emph{f} states can be
written as
\begin{eqnarray}
\label{wf_init} \Psi_{JM}(\mathbf{x_{1}},\mathbf{x_{2}}) =
{A_N}\sum_{m_{1}m_{2}}\sum_{\mathcal{P}}(-1)^{\mathcal{P}}\mathcal{P}C_{j_{1}m_{1},j_{2}m_{2}}^{JM}
\psi_{j_{1}m_{1}}(\mathbf{x_{1}})\psi_{j_{2}m_{2}}(\mathbf{x_{2}}),
\end{eqnarray}
where $\psi_{jm}(\mathbf{x})$ is the
one-electron Dirac wave function,
 $A_N = 1/2$ for equivalent electrons and $A_N = 1/\sqrt{2}$
for  nonequivalent electrons.
To account for the weak interaction we have to modify the wave
function of the final state by admixing the close opposite-parity
state:
\begin{eqnarray}
\label{wf_pnc1} |2^1S_0\rangle \rightarrow |2^1S_0\rangle
+\frac{\langle 2^3P_0|H_W(1)+H_W(2)|2^1S_0\rangle}{E_{2^1S_0}-
E_{2^3P_0}}|2^3P_0\rangle\,,\\
|2^3P_0\rangle \rightarrow |2^3P_0\rangle +\frac{\langle
2^1S_0|H_W(1)+H_W(2)|2^3P_0\rangle}{E_{2^3P_0}-
E_{2^1S_0}}|2^1S_0\rangle\,. \label{wf_pnc2}
\end{eqnarray}
Here we have introduced the spin-independent part of the effective
nuclear weak-interaction Hamiltonian \cite{khr91}
\begin{eqnarray}
\label{h_w_m} \label{h_w} H_W=-(G_F/\sqrt{8})Q_W
\rho_{N}(r)\gamma_5,
\end{eqnarray}
where $G_F$ is the Fermi constant,  $Q_W \approx -N + Z(1-4{\rm
sin}^2\theta_W) $ is the weak charge of the nucleus, $\rho_{N}$ is
the nuclear weak-charge density normalized to unity, and $\gamma_5$
is the Dirac matrix. For convenience, we rewrite formulas
(\ref{wf_pnc1})-(\ref{wf_pnc2}) as
\begin{eqnarray}
\label{wf_pnc1p}
|2^1S_0\rangle \rightarrow |2^1S_0\rangle +i\xi|2^3P_0\rangle\,,\\
|2^3P_0\rangle \rightarrow |2^3P_0\rangle +i\xi|2^1S_0\rangle\,,
\label{wf_pnc2p}
\end{eqnarray}
or in general
\begin{eqnarray}
\label{wf_pnc1p} |\Psi_{f}\rangle \rightarrow |\Psi_{f}\rangle
+i\xi|\Psi_{\tilde{f}}\rangle\,,  \label{wf_pncm}
\end{eqnarray}
where the parameter $\xi$ is given by
\begin{eqnarray}
\label{wf_pnc1p} \label{xi} \xi =
\frac{G_F}{2\sqrt{2}}\frac{Q_W}{E_{2^1S_0}-E_{2^3P_0}}
\int_{0}^{\infty}dr\,r^2
\rho_N(r)[g_{2p_{1/2}}f_{2s}-f_{2p_{1/2}}g_{2s}]
\end{eqnarray}
with the large and small radial components of the Dirac wave
function  defined by
\begin{eqnarray} \psi_{n\kappa m}({\bf r})=
\left(\begin{array}{c}
g_{n\kappa}(r)\Omega_{\kappa m}({\bf n})\\
if_{n\kappa}(r)\Omega_{-\kappa m}({\bf n})
\end{array}\right)\;.
\end{eqnarray}
With the PNC correction, the differential cross section is given by
\begin{eqnarray} \label{cross_pnc}
\sigma= \frac{(2\pi)^{4}}{{v_{i}}}\mathbf{k_f^2} \sum_{\beps_f}
\left\{\left|\sum_{M_d}\tau_{\gamma_f,f;d}
\frac{1}{E_i-E_d+i\Gamma_d/2}
\langle\Psi_d|I|\Psi_i\rangle+\tau_{\gamma_f,f;i}
\right|^{2}\nonumber\right.\\
\left.+2\Re\left[i\xi\left(\sum_{M_d}
\tau_{\gamma_f,f;d}\frac{1}{E_i-E_d+i\Gamma_d/2}
\langle\Psi_d|I|\Psi_i\rangle+\tau_{\gamma_f,f;i}
\right)\nonumber\right.\right.\\
\left.\left.\times\left(\sum_{M_d'}
\tau^{*}_{\gamma_f,\tilde{f};d} \frac{1}{E_i-E_d-i\Gamma_d/2}
\langle\Psi_d|I|\Psi_i\rangle^{*}+\tau^{*}_{\gamma_f,\tilde{f};i}
\right)\right]\right\}.
\end{eqnarray}

\begin{figure}
\includegraphics[width=\columnwidth]{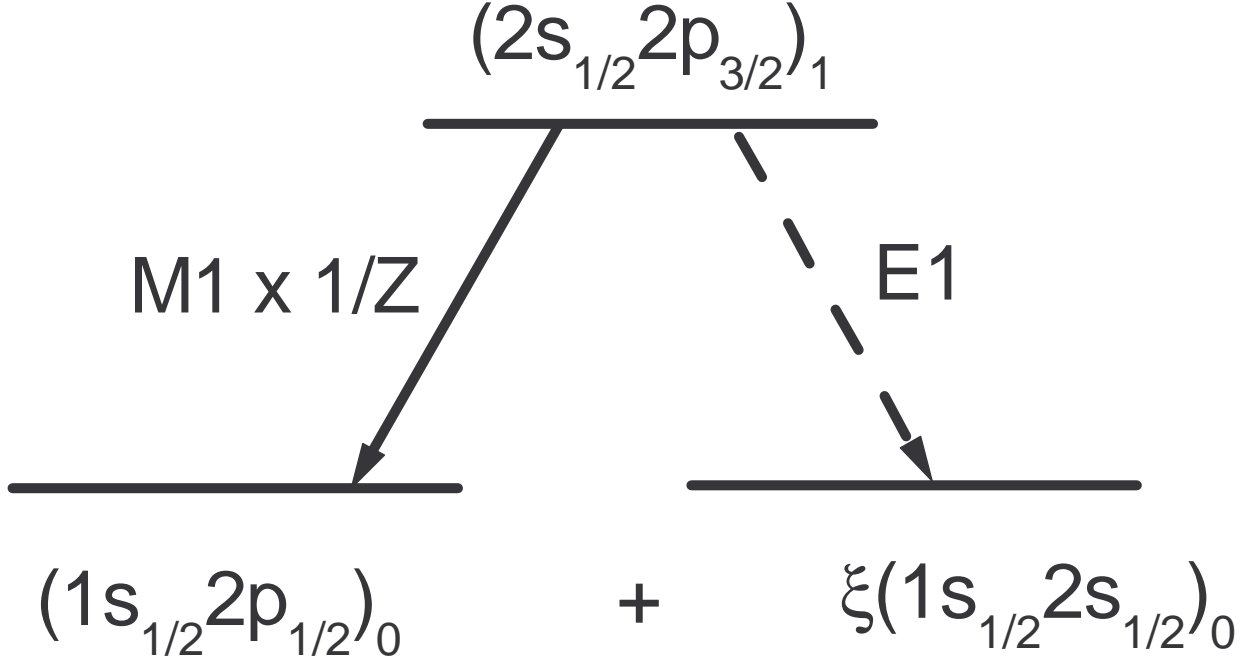}\caption{The decay scheme of the $|(2s_{1/2}2p_{3/2})_{1}\rangle$
state into the $|2^3P_0\rangle +\xi |2^1S_0\rangle$ state. The
main channel  is the two-electron M1 transition, which is
suppressed by a factor $1/Z$, while the PNC-mixing channel is the
one-electron E1 transition.} \label{fig0}
\end{figure}

To a good approximation, the RR amplitudes are calculated by formulas
\begin{eqnarray}\label{trr1}
\tau_{\gamma_f,f;i}
=-\langle\Psi_{f}|{R}^{\dagger}(1)+{R}^{\dagger}(2)|\Psi_i\rangle\,,\\
\tau_{\gamma_f,\tilde{f};i}
=-\langle\Psi_{\tilde{f}}|{R}^{\dagger}(1)+{R}^{\dagger}(2)|\Psi_i\rangle\,,
\label{trr2}
\end{eqnarray}
where
${R} =e\alpha_{\mu}A^{\mu}=-e\bm{\alpha}\cdot {\bf A}$ is the
transition operator acting on the electron variables, and
\begin{eqnarray}
{\bf A}(\bfx) = \frac{\beps \exp{(i\bfk_f\cdot \bfx)}}{\sqrt{2k_f^0(2\pi)^3}}\,
\label{photon}
    \end{eqnarray}
is the wave function of the emitted photon. Calculations of the RR transition
amplitude including the PNC effect were considered in detail in Ref. \cite{mai09}.

To enhance the PNC effect on the dielectronic recombination, one should consider
the resonance with a state $d$ which decays to the final $f$ state via the M1 transition
while the admixture of the $\tilde{f}$ state due to the weak interaction
 enables the E1 transition. As the simplest case, one can choose
the state $d=(2s_{1/2} 2p_{1/2})_1$, which decays into
the state $f =  2^3P_0$ via the one-electron
M1 transition, whereas the PNC mixing transition is the one-electron E1 transition.
In this case the $\tau_{\gamma_f,f;d}$ and $\tau_{\gamma_f,\tilde{f};d}$
amplitudes can be calculated within the lowest-order approximation by
formulas which are similar to (\ref{trr1})-(\ref{trr2}),
\begin{eqnarray}
\tau_{\gamma_f,f;d} = -\langle\Psi_f|{R^{\dagger}}(1)+{R^{\dagger}}(2)|\Psi_d\rangle\,, \\
\tau_{\gamma_f,\tilde{f};d} = -\langle\Psi_{\tilde{f}}|{R^{\dagger}}(1)+{R^{\dagger}}(2)|\Psi_d\rangle\,.
\end{eqnarray}
A bigger enhancement of the PNC effect on
the dielectronic recombination should occur
 in a situation where the doubly excited state $d$ decays
into the final $f$ state via the two-electron M1 transition, which
is additionally suppressed by a factor $1/Z$ compared to the one-electron
M1 transition, while the PNC mixing transition is the one-electron E1
transition.
As an example, we can consider $d = (2s_{1/2}2p_{3/2})_{1}$.
The related decay scheme
is shown in Fig.~\ref{fig0}, where
the solid line displays the main channel that is
the two-electron M1 transition
  and the dashed line  indicates the PNC-mixing channel that is
the one-electron E1 transition.
Since the E1 transition is very strong compared to the
M1$\times1/Z$ transition, the admixture of the $2^1S_0$
state to the $2^3P_0$ state due to the weak interaction
should be most pronounced in the DR contribution.

The two-electron transition amplitude can be evaluated by
perturbation theory \cite{sha02,ind04}.
Let the two-electron states
\emph{d} and \emph{f} are defined by sets of one-electron states
($d_1, d_2$) and ($f_1, f_2$), respectively.
For the case under consideration
(all four states $d_{1}$, $d_{2}$, $f_{1}$ and $f_{2}$ are different)
the two-electron transition amplitude is given by (for more details
see Ref. \cite{ind04})
\begin{eqnarray}
\tau_{\gamma_f,f;d} =
C_N\sum_{m_{d_{1}},m_{d_{2}}}C_{j_{d_{1}}m_{d_{1}},j_{d_{2}}m_{d_{2}}}^{J_d
M_d}\sum_{m_{f_{1}},m_{f_{2}}}C_{j_{f_{1}}m_{f_{1}},j_{f_{2}}m_{f_{2}}}^{J_f
M_f}(\tau^{a}_{\gamma_f,f;d} +\tau^{b}_{\gamma_f,f;d}),
\end{eqnarray}
where
\begin{eqnarray}
\tau^{a}_{\gamma_f,f;d} =
-\sum_{\mathcal{P}}(-1)^{\mathcal{P}}\left\{\sum_{n}\langle\mathcal{P}f_{1}|e\alpha_{\mu}A^{\mu*}|n
\rangle\frac{1}{E_{d}^{(0)}-\varepsilon_{\mathcal{P}f_{2}}-\varepsilon_{n}}\langle
n\mathcal{P}f_{2}|I(\varepsilon_{\mathcal{P}f_{2}}-\varepsilon_{d_{2}})|d_{1}d_{2}\rangle
\nonumber\right.\\ \left.+
\sum_{n}\langle\mathcal{P}f_{2}|e\alpha_{\mu}A^{\mu*}|n
\rangle\frac{1}{E_{d}^{(0)}-\varepsilon_{\mathcal{P}f_{1}}-\varepsilon_{n}}\langle
\mathcal{P}f_{1}n|I(\varepsilon_{\mathcal{P}f_{1}}-\varepsilon_{d_{1}})|d_{1}d_{2}\rangle\right\},
\end{eqnarray}
\begin{eqnarray}
\tau^{b}_{\gamma_f,f;d} =
-\sum_{\mathcal{P}}(-1)^{\mathcal{P}}\left\{\sum_{n}\langle
\mathcal{P}f_{1}\mathcal{P}f_{2}|I(\varepsilon_{\mathcal{P}f_{2}}-\varepsilon_{d_{2}})|nd_{2}\rangle
\frac{1}{E_{f}^{(0)}-\varepsilon_{d_{2}}-\varepsilon_{n}}
\langle n|e\alpha_{\mu}A^{\mu*}|d_{1} \rangle \nonumber\right.\\
\left.+ \sum_{n}\langle
\mathcal{P}f_{1}\mathcal{P}f_{2}|I(\varepsilon_{\mathcal{P}f_{1}}-\varepsilon_{d_{1}})|d_{1}n\rangle
\frac{1}{E_{f}^{(0)}-\varepsilon_{d_{1}}-\varepsilon_{n}} \langle
n|e\alpha_{\mu}A^{\mu*}|d_{2} \rangle\right\},
\end{eqnarray}
$C_N$ is the normalization factor which is equal to 1 in case
of non-equivalent electrons in both initial and final states,
$E_{f}^{(0)} = \varepsilon_{f_{1}}+\varepsilon_{f_{2}}$,
$E_{d}^{(0)} = \varepsilon_{d_{1}}+\varepsilon_{d_{2}}$,
and $\varepsilon_{n}$ is the one-electron Dirac energy.

So far we have assumed that the intermediate and final states are
well isolated single levels. This approximation, being valid  for
the final states, can be incorrect for the intermediate states
which have rather large widths. To account for  quasidegeneracy of
the intermediate states within the rigorous QED approach, one can
use the methods described in  Refs. \cite{sha02,and08}. For our
purposes, however, it is sufficient  to use a simple replacement
of
 the expression (\ref{cross_pnc}) by
one which contains a sum over all close-lying intermediate
states and account for the mixing of the states having
the same symmetry.

%
%

\section{Numerical results and discussion}

The most promising situation for observing the PNC effect in the
process under consideration occurs for gadolinium ($Z = 64$) and
thorium ($Z = 90$). The energy difference between the levels
$2^{1}S_{0}$ and $2^{3}P_{0}$ amounts to  -0.023(74) for
gadolinium and to  0.44(40) eV for thorium
\cite{art05,kozh08b,sha10}. As in our previous papers
\cite{mai09,sha10}, to estimate the PNC effect we use 0.44 eV and
0.074 eV for the  $2^{3}P_{0}-2^{1}S_{0}$ energy difference in
cases of Th and Gd, respectively.

Let us consider the requirement that should be imposed on the luminosity $L$,
provided the PNC effect is measured to a relative accuracy $\eta$. Denoting
by $\sigma_+$ and $\sigma_-$ the differential cross sections for the
positive and negative spin projection of the incident electron
onto the electron momentum, one can derive \cite{gri05,mai09}
\begin{eqnarray}
L>L_{0}=\frac{\sigma_+ + \sigma_-+2\sigma_{\rm
b}}{(\sigma_+-\sigma_-)^2\eta^2 T} \,. \label{lum}
\end{eqnarray}
where $\sigma_{\rm b}$ is the background magnitude  and $T$ is the
acquisition time. In our calculations we  set $T$ equal to two weeks
and  neglect the background signal.
We have studied resonance recombination processes which
correspond to the following DR channels
(the admixture due to the weak interaction is implied):

1) $\bar{e} + 1s \rightarrow (2s_{1/2} 2p_{1/2})_1 \rightarrow
(1s_{1/2}2p_{1/2})_0 + \gamma$

2)  $\bar{e} + 1s \rightarrow (2s_{1/2} 2p_{3/2})_1 \rightarrow
(1s_{1/2}2p_{1/2})_0 + \gamma$

3)  $\bar{e} + 1s \rightarrow (2p_{1/2} 2p_{3/2})_1 \rightarrow
(1s_{1/2}2s_{1/2})_0 + \gamma$





4) $\bar{e} + 1s \rightarrow (2s_{1/2} 3p_{1/2})_1 \rightarrow
(1s_{1/2}2p_{1/2})_0 + \gamma$

5) $\bar{e} + 1s\rightarrow (2s_{1/2} 3p_{3/2})_1 \rightarrow
(1s_{1/2}2p_{1/2})_0 + \gamma$ \\
The first process is the only one in which the M1 transition
from the resonance state is
not suppressed by a factor 1/Z. All other processes include
the two-electron M1
transition to the final state
while the PNC-mixing channel is the one-electron E1
transition.
For all  processes we evaluated the differential cross
section (\ref{cross_pnc})
as a function of the photon emission
angle ($\theta$) with respect to the incident electron momentum.
The calculations of the Dirac wave functions that enter the formulas
were performed using the RADIAL package \cite{sal95} and the
dual-kinetic-balance basis set method  \cite{sha04}  with the basis functions
constructed from B-splines \cite{sap96}.

\begin{figure}
\vspace*{12pt}
\includegraphics[width=\columnwidth]{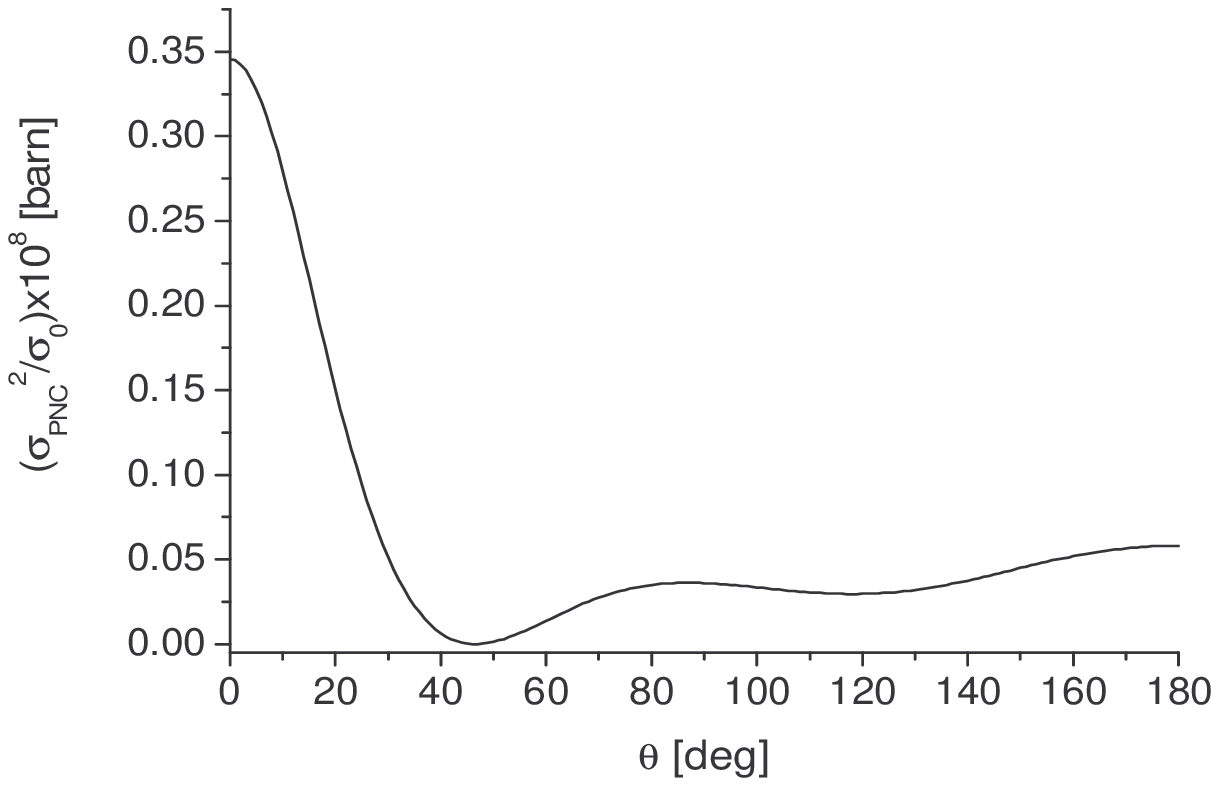}
\caption{The value $\sigma_{\rm PNC}^2/\sigma_0 \sim 1/L_0$ as a
function of photon emission angle
 $\theta$ for the resonance recombination into the
$2^3P_0$ state of He-like thorium with intermediate state
$(2s_{1/2}2p_{1/2})_1$, 1-st process.} \label{fig1}
\end{figure}

\begin{figure}
\vspace*{12pt}
\includegraphics[width=\columnwidth]{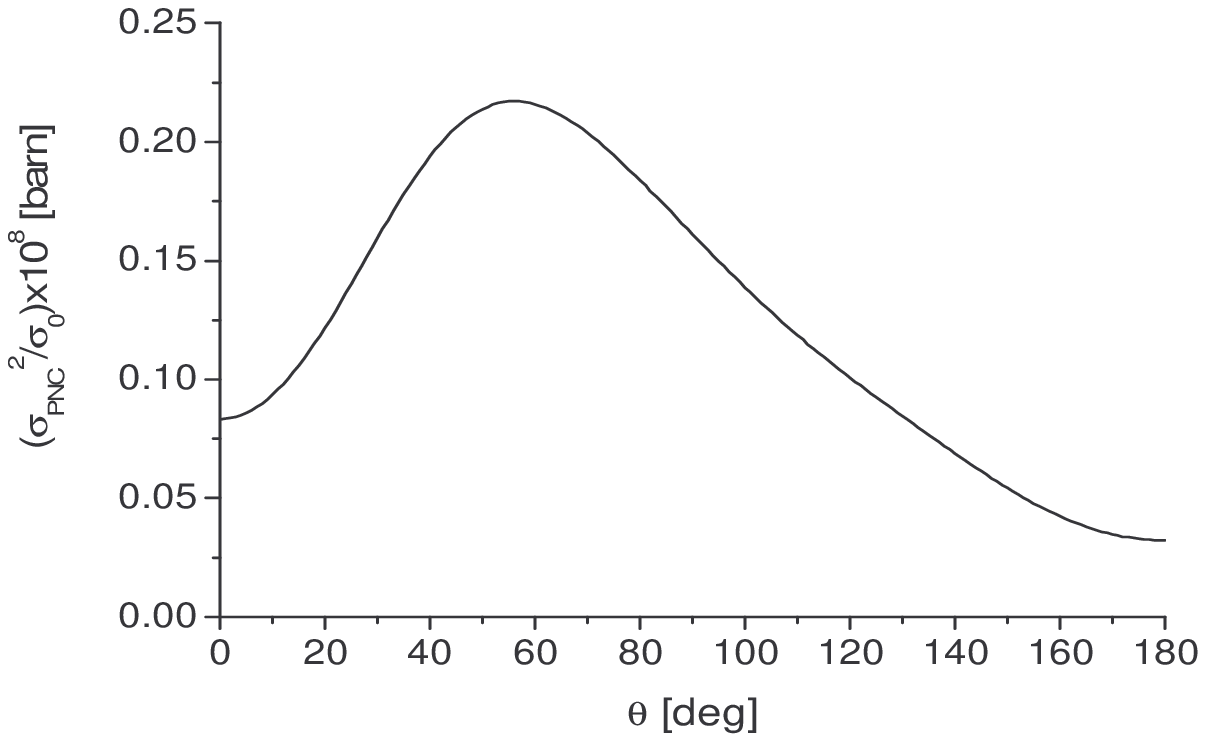}
\caption{The value $\sigma_{\rm PNC}^2/\sigma_0 \sim 1/L_0$ as a
function of
photon emission angle
 $\theta$ for the resonance recombination into the
$2^3P_0$ state of He-like thorium with intermediate state
$(2s_{1/2}2p_{3/2})_1$, 2-nd process.} \label{fig2}
\end{figure}

\begin{figure}
\vspace*{12pt}
\includegraphics[width=\columnwidth]{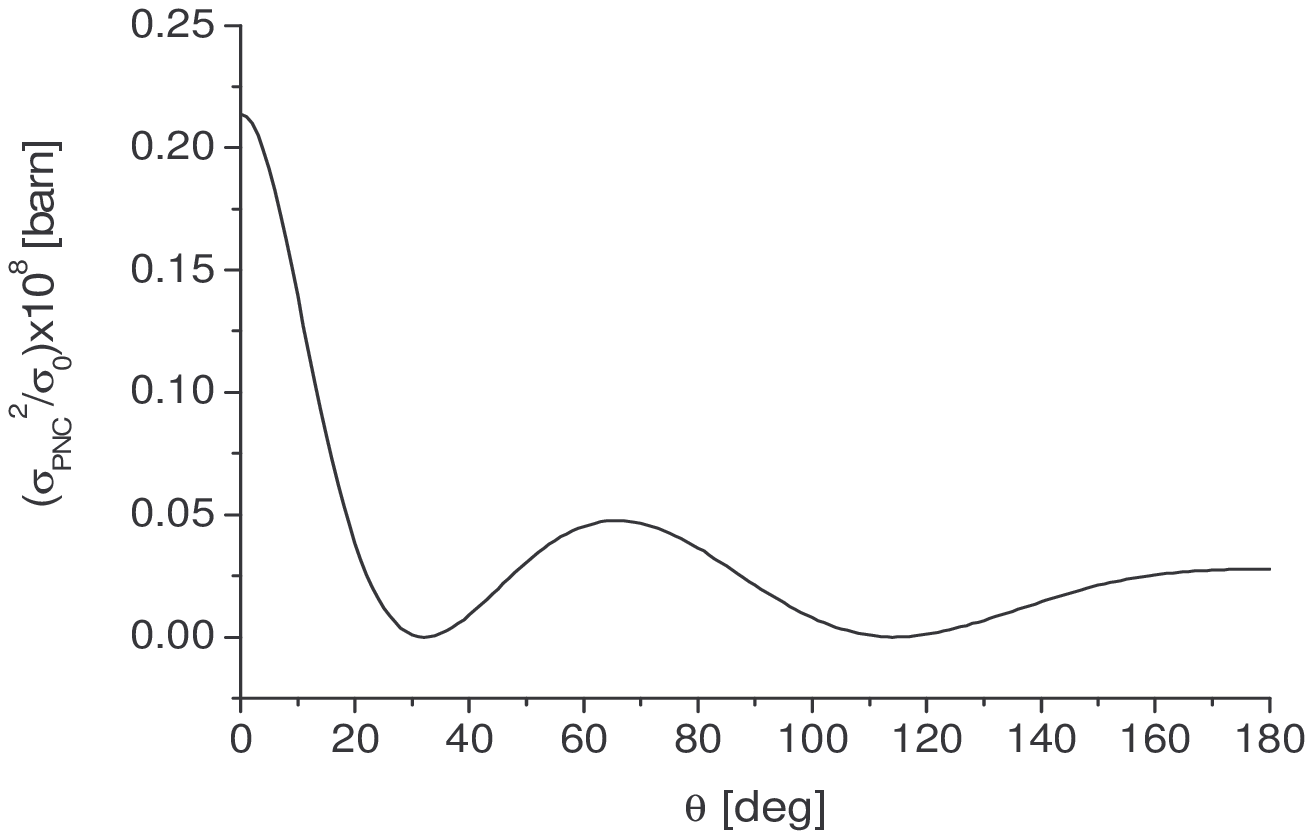}
\caption{The value $\sigma_{\rm PNC}^2/\sigma_0 \sim 1/L_0$ as a
function of  photon emission angle
$\theta$ for the resonance recombination into the
$2^3P_0$ state of He-like thorium with intermediate state
$(2s_{1/2} 3p_{1/2})_1$, 4-th process.} \label{fig3}
\end{figure}

Table I presents numerical results for the differential cross section
in case of thorium at the angles $\theta$
corresponding to the minimum values of the luminosity. Table II
presents the related results for gadolinium. We denote the cross
section without the PNC effect as $\sigma_0 = (\sigma_+
+\sigma_-)/2$ and the PNC contribution as $\sigma_{\rm PNC} =
(\sigma_+ -\sigma_-)/2$. $N$ is the process number in the list
above.
As can be seen from the tables, the most favourable are
the 1-st, 2-nd, and 4-th
processes. In all these cases the final
state is $|2^3P_0\rangle$ with an admixture of  $|2^1S_0\rangle$
 due to the weak interaction. In
Figs.~\ref{fig1},~\ref{fig2}, and ~\ref{fig3}
we display the values $\sigma_{\rm PNC}^2/\sigma_0 \sim 1/L_0$ as
functions of $\theta$ for 1-st, 2-nd, and 4-th processes.
According to the tables,   the PNC asymmetry
of the cross section does exceed 0.01\%.
It should be stressed, however, that the PNC asymmetry on some of
the DR contributions, taken separately, is very large and for the
resonance DR into the  $(2p_{1/2} 2p_{3/2})_1$ state amounts to about 12\%
in case of thorium. This  is extremely large value
for atomic PNC  effects.  But, unfortunately, this large value is strongly
masked by the RR and non-resonance DR contributions.
This fact, together with current restrictions on the experimental resolution
(see the related discussion in Ref. \cite{mai09}),
make practical realization of such an experiment  rather problematic.
We think, however, that the calculations performed will help us to search
for more realistic scenarios to observe the  PNC effect in resonance
scattering processes
with heavy few-electron ions.

\begin{table} \caption{Differential cross section for resonance
recombination of a polarized electron with H-like thorium at the
photon emission angle $\theta$ corresponding
to the minimum value of the luminosity $L_0$ which is defined by
Eq. (\ref{lum}) at $T$ = 2 weeks. $\sigma_0$ is the cross section
without the PNC effect and $\sigma_{\rm PNC}$ is the PNC
contribution. Results are presented in ascending order of the
luminosity $L_0$, $N$ is the process number defined in the text.
 } \vspace{0.2cm}
\label{tab2}
\begin{tabular} {cccccc}
\hline  $N$ &$\varepsilon_{i}$ [keV] &$\theta$ [deg]
& $L_0$ [cm$^{-2}$s$^{-1}$] & $\sigma_0$ [barn] & $\sigma_{\rm PNC}$ [barn]\\
\hline
1 & 60.91 & 0 & 1.2$\times 10^{30}$ & 1.076 & 6.1$\times 10^{-5}$  \hspace*{0.115cm} \\
2 & 65.04 & 56 & 1.9$\times 10^{30}$ & 1.247 & -5.2$\times 10^{-5}$  \hspace*{0.115cm} \\
4 & 79.21 & 0 & 1.9$\times 10^{30}$ & 0.429 & 3.0$\times 10^{-5}$ \hspace*{0.115cm} \\
5 & 80.41 & 53 & 3.2$\times 10^{30}$ & 0.779 & -3.2$\times 10^{-5}$ \hspace*{0.115cm} \\
3 & 64.96 & 43 & 2.0$\times 10^{31}$ & 0.287 & 7.7$\times 10^{-6}$  \hspace*{0.115cm} \\
\hline
\end{tabular}
\end{table}

\begin{table} \caption{Differential cross section for resonance
recombination of a polarized electron with H-like gadolinium at the
photon emission angle $\theta$ corresponding
to the minimum value of the luminosity $L_0$ which is defined by
Eq. (\ref{lum}) at $T$ = 2 weeks. $\sigma_0$ is the cross section
without the PNC effect and $\sigma_{\rm PNC}$ is the PNC
contribution. Results are presented in ascending order of the
luminosity $L_0$, $N$ is the process number defined in the text.
 } \vspace{0.2cm}
\label{tab2}
\begin{tabular} {cccccc}
\hline  $N$ &$\varepsilon_{i}$ [keV] &$\theta$ [deg]
& $L_0$ [cm$^{-2}$s$^{-1}$] & $\sigma_0$ [barn] & $\sigma_{\rm PNC}$ [barn]\\
\hline
2 & 30.27 & 66 & 4.3$\times 10^{30}$ & 0.996 & 3.1$\times 10^{-5}$  \hspace*{0.115cm} \\
5 & 37.93 & 71 & 7.5$\times 10^{30}$ & 0.568 & 1.8$\times 10^{-5}$ \hspace*{0.115cm} \\
1 & 29.36 & 0 & 7.7$\times 10^{30}$ & 0.467 & -1.6$\times 10^{-5}$ \hspace*{0.115cm} \\
4 & 37.67 & 0 & 2.1$\times 10^{31}$ & 0.218 & -6.6$\times 10^{-6}$ \hspace*{0.115cm} \\
3 & 30.24 & 40 & 1.9$\times 10^{32}$ & 0.331 & -2.7$\times 10^{-6}$  \hspace*{0.115cm} \\

\hline
\end{tabular}
\end{table}

In summary, we have studied the PNC effect on the cross section
of  resonance recombination of polarized electrons  with H-like
thorium and gadolinium, where the PNC effect is strongly enhanced due
quasidegeneracy of the opposite-parity $2^1S_0$ and  $2^3P_0$ states.
The calculations were performed for different
intermediate doubly excited states, which can decay into one of the
$2^1S_0$ or $2^3P_0$ states  via one-photon  emission.
It was found that the most promising situation occurs
when the incident electron energy is chosen to be in resonance
with the  $(2s_{1/2}2p_{1/2})_{1}$ state. We hope that this work
will stimulate further  efforts for  studying
 the PNC effects in resonance scattering processes
with heavy ions.

\acknowledgments
{This work was supported by DFG (Grants No. PL 254/7-1 and VO 1707/1-1),
by RFBR (Grant No. 10-02-00450), by GSI, by DAAD,
 by the Ministry of Education and Science of Russian Federation
(Program ``Scientific and pedagogical specialists for innovative Russia'', Grant No. P1334).
The work of A.V.M.  and  V.A.Z. was also supported by the ``Dynasty'' foundation.}

\end{document}